\documentstyle[aps]{revtex}

\begin{document}
\draft
\title{Angular momentum conservation law in Einstein-Cartan space-time}
\author{Ying Jiang\thanks{%
E-mail: yjiang@itp.ac.cn}}
\address{CCAST (World Laboratory), Box 8730, Beijing 100080, P. \ R. China}
\address{Institute of Theoretical Physics, Chinese Academy of Sciences, P. O. Box
2735, Beijing 100080, P. R. China\thanks{%
mailing address}}
\maketitle

\begin{abstract}
In the light of the local Lorentz transformations and the general Noether
theorem, a new formulate of the general covariant angular momentum
conservation law in Einstein-Cartan gravitation theory is obtained, which
overcomes the critical difficulty in the other formulates that the
conservation law depended on the coordinative choice.
\end{abstract}


\section{Introduction}

Conservation law of energy-momentum and angular momentum have been of
fundamental interest in gravitational physics\cite{89}. Using the vierbein
representation of general relativity, Duan {\it et al} obtained a general
covariant conservation law of energy-momentum which overcomes the
difficulties of other expressions\cite{43}. This conservation law gives the
correct quadrupole radiation formula of energy which is in good agreement
with the analysis of the gravitational damping for the pulsar PSR1916-13\cite
{39}. Also, from the same point of view, Duan and Feng\cite{27} proposed a
general covariant conservation law of angular momentum in Riemannian
space-time which does not suffer from the flaws of the others\cite{53,4,16}.

On the other hand, though the Einstein theory of general relativity and
gravitation has succeeded in many respects, there is an essential difficulty
in this theory: we could not get a successful renormalized quantum gravity
theory\cite{96}. In order to find renormalized theory, many physicists\cite
{69,85} have studied this problem in its more general aspects, i.e.
extending Einstein's theory to Einstein-Cartan theory, which includes
torsion tensor\cite{63}.

As is well known, torsion is a slight modification of the Einstein's theory
of relativity\cite{23}, but is a generalization that appears to be necessary
when one tries to conciliate general relativity with quantum theory. Like
opening a Pandora's box , many works have been done in this region\cite
{104,77}. Today, general relativity with non-zero torsion is a major
contender for a realistive generalization of the theory of gravitation.

About two decades ago, Helh\cite{63} gave, in Einstein-Cartan theory, an
expression of the angular momentum conservation law which was worked out
from Noether' theorem, but in that expression, all quantities carried
Riemannian indices and the total angular momentum depended on the
coordinative choice which is not an observable quantity. Some physicists\cite
{69,60} investigated the same problem from different viewpoints and
presented other expressions of conservation law which is not general
covariant, hence this theory cannot be said to be very satisfactory.

Several years ago, the general covariant energy-momentum
conservation law in general space-time has been discussed
successfully by Duan {\it et al}\cite {38}. In this paper, we will
discuss the angular momentum conservation law in Einstein-Cartan
theory via the vierbein representation. General Relativity without
vierbein is like a boat without a jib--without these vital
ingredients the going is slow and progress inhibited.
Consequently, vierbein have grown to be an indispensable tool in
many aspects of general relativity. More important, it is relevant
to the physical observability. Based on the Einstein's observable
time and space interval, we take the local point of view that any
measurement in physics is performed in the local flat reference
system whose existence is guaranteed by the equivalence principle,
i.e. an observable object must carries, instead of the indices of
the space-time coordinates, the indices of internal space. Thus,
we draw the support from vierbein not only for mathematical
reason, but also because of physical measurement consideration.

\section{Conservation law in general case}

The conservation law is one of the important essential problems in
gravitational theory. It is due to the invariant of Lagrangian corresponding
to some transformation. In order to study the general covariant angular
momentum conservation law, it is necessary to discuss conservation law by
the Noether theorem in general case.

The action of a system is
\begin{equation}  \label{1}
I=\int_{M} {\cal L} (\phi ^A, \phi ^A _{,\mu}) d^4x,
\end{equation}
where $\phi ^A$ are independent variable with general index $A$, $\phi ^A
_{,\mu}=\partial _{\mu} \phi ^A$. If the action is invariant under the
infinitesimal transformation
\begin{equation}  \label{2}
x^{^{\prime}\mu}=x^{\mu}+\delta x^{\mu},
\end{equation}
\begin{equation}  \label{3}
\phi ^{^{\prime}A}(x ^{^{\prime}})=\phi ^A(x) + \delta \phi ^A (x),
\end{equation}
and $\delta \phi ^A$ is zero on the boundary of the four-dimensional volume $%
M$, then we can prove that there is the relation
\begin{equation}  \label{4}
\frac {\partial}{\partial x^{\mu}}({\cal L} \delta x ^{\mu} + \frac{\partial
{\cal L}}{\partial \phi ^{A} _{,\mu}}) +[{\cal L}]_{\phi ^A}\delta _0 \phi
^A=0
\end{equation}
where $[{\cal L}] _{\phi ^A}$ is
\begin{equation}  \label{5}
[{\cal L}] _{\phi ^A}=\frac {\partial {\cal L}}{\partial \phi ^A} - \partial
_{\mu}(\frac {\partial {\cal L}}{\partial \phi ^A _{, \mu}}),
\end{equation}
and $\delta _0 \phi ^A$ is the Lie derivative of $\phi ^A$
\begin{equation}  \label{6}
\delta _0 \phi ^A= \delta \phi ^A (x) -\phi ^A _{,\mu} \delta x ^{\mu}.
\end{equation}

If ${\cal L}$ is the total Lagrangian density of the system, there is $[%
{\cal L}] _{\phi ^A}=0$, the field equation of $\phi ^A$ with respect to $%
\delta I =0$. From the above equation, we know that there is a conservation
equation corresponding to the above transformations
\begin{equation}  \label{7}
\frac {\partial}{\partial x ^{\mu}}({\cal L} \delta x^{\mu} + \frac {%
\partial {\cal L}}{\partial \phi ^A _{,\mu}} \delta _0 \phi ^A)=0
\end{equation}
This is just the conservation law in general case. It must be pointed out
that if ${\cal L}$ is not the total Lagrangian density of the system, then
as long as the action of ${\cal L}$ remains invariant under these
transformations, (\ref{4}) is still tenable. But (\ref{7}) is not admissible
now due to $[{\cal L}]_{\phi ^A} \neq0$.

In gravitation theory with the vierbein as element fields, we can separate $%
\phi ^A$ as $\phi ^A =(e ^a _{\mu}, \psi ^B)$, where $e ^a _{\mu}$ is the
vierbein field and $\psi ^B$ is an arbitrary tensor under general coordinate
transformation. When $\psi ^B$ is $\psi ^{\mu _1 \mu _2 \cdots \mu _k}$, we
can always scalarize it by
\[
\psi ^{a _1 a _2 \cdots a _k} = e ^{a _1} _{\mu _1} e ^{a _2} _{\mu _2}
\cdots e ^{a _k} _{\mu _k} \psi ^{\mu _1 \mu _2 \cdots \mu _k},
\]
so we can take $\psi ^B$ as a scalar field under general coordinate
transformations. In later discussion we can simplify the equations by such a
choice.

\section{General Covariant conservation law of Angular momentum in
Einstein-Cartan theory}

As is well known, in Einstein-Cartan theory, the total action of the
gravity-matter system is expressed by
\begin{equation}
I=\int_{M}{\cal L}d^{4}x=\int_{M}({\cal L}_{g}+{\cal L}_{m})d^{4}x,
\end{equation}
\begin{equation}
{\cal L}_{g}=\frac{c^{4}}{16\pi G}\sqrt{-g}R
\end{equation}
${\cal L}_{g}$ is the gravitational Lagrangian density, $R$ is the scalar
curvature of the Riemann-Cartan space-time. The matter part Lagrangian
density ${\cal L}_{m}$ take the form ${\cal L}_{m}={\cal L}_{m}(\phi
^{A},D_{\mu }\phi ^{A})$, where the matter field $\phi ^{A}$ belongs to some
representation of Lorentz group whose generators are $I_{ab}$ $(a,b=0,1,2,3)$
and $I_{ab}=-I_{ba}$, $D_{\mu }$ is the covariant derivative operator of $%
\phi ^{A}$
\begin{equation}
D_{\mu }\phi ^{A}=\partial _{\mu }\phi ^{A}-\frac{1}{2}\omega _{\mu
ab}(I_{ab})_{B}^{A}\phi ^{B}.
\end{equation}
where $\omega _{\mu ab}$ is the spin connection.

As in Einstein-Cartan theory, the affine connection $\Gamma _{\mu \nu
}^{\lambda }$ is not symmetry in $\mu $ and $\nu $, i.e. there exists
non-zero torsion tensor
\begin{equation}
T_{\mu \nu }^{\lambda }=\Gamma _{\mu \nu }^{\lambda }-\Gamma _{\nu \mu
}^{\lambda }.
\end{equation}
It is well known, for vierbein field $e_{\mu }^{a}$, the total covariant
derivative is equal to zero, i.e.
\begin{equation}
{\cal D}_{\mu }e_{\nu }^{a}\equiv \partial _{\mu }e_{\nu }^{a}-\omega _{\mu
ab}e_{\nu }^{b}-\Gamma _{\mu \nu }^{\lambda }e_{\lambda }^{b}=0.
\end{equation}
From this formula, we can obtain another expression of torsion tensor
\begin{equation}
T_{\mu \nu }^{a}=T_{\mu \nu }^{\lambda }e_{\lambda }^{a}=D_{\mu }e_{\nu
}^{a}-D_{\nu }e_{\mu }^{a},
\end{equation}
where $D_{\mu }e_{\nu }^{a}=\partial _{\mu }e_{\nu }^{a}-\omega _{\mu
}^{ab}e_{\nu }^{b}$. In fact, this formula is just the Cartan structure
equation.

The scalar curvature of the Riemann-Cartan space-time is expressed
by
\begin{eqnarray}
R &=& e_{a}^{\mu }e_{b}^{\nu }\partial _{\mu }\omega _{\nu
ab}-e_{a}^{\mu
}e_{b}^{\nu }\partial _{\nu }\omega _{\mu ab}+\bar{\omega}_{bac}\bar{\omega}%
_{acb}+\bar{\omega}_{b}\bar{\omega}_{b} \nonumber \\
&+& \bar{\omega}_{acb}T_{acb}-2\bar{\omega}_{b}T_{b}+T_{b}T_{b}+\frac{1}{2}%
T_{cba}T_{bac}+\frac{1}{4}T_{acb}T_{abc}.
\end{eqnarray}
where $\omega _{abc}=e_{a}^{\mu }\omega _{\mu bc},\;\omega
_{a}=\omega
_{bab} $ are the representation of spin connection, $\bar{\omega}_{a}=\bar{%
\omega}_{bab}=e_{b}^{\mu }(\partial _{\mu }e_{a}^{\nu }+\{_{\mu \sigma
}^{\nu }\}e_{a}^{\sigma })e_{\nu b}$ in which $\{_{\mu \sigma }^{\nu }\}$ is
the Christoffel symbol. By tedious calculation, the above formulation allows
us to obtain the identity
\begin{equation}
{\cal L}_{g}=\frac{c^{4}}{16\pi G}\sqrt{-g}R=\frac{c^{4}}{8\pi G}{\cal L}%
_{\partial }+{\cal L}_{\omega }
\end{equation}
\begin{equation}
{\cal L}_{\partial }=\partial _{\mu }(\sqrt{-g}e_{a}^{\mu }\omega _{a})
\label{total}
\end{equation}
\begin{equation}
{\cal L}_{\omega }=\frac{c^{4}}{16\pi G}\sqrt{-g}[\omega
_{bac}\omega _{acb}+\omega _{a}\omega _{a}-2\bar{\omega}_{a}\omega
_{a}-2e_{a}^{\mu }\partial _{\mu }(e_{b}^{\nu })e_{\nu c}\omega
_{cab}]
\end{equation}

It is well known that in deriving the general covariant
conservation law of energy momentum in general relativity, the
general displacement transformation, which is a generalization of
the displacement transformation in the Minkowski space-time, was
used\cite{38}. In the local Lorentz reference frame, the general
displacement transformation takes the same form as that in the
Minkowski space-time. This implies that general covariant
conservation laws are corresponding to the invariance of the
action under local transformations. We may conjecture that since
the conservation law for angular momentum in special relativity
corresponds to the invariance of the action under the Lorentz
transformation, the general covariant conservation law of angular
momentum in general relativity may be obtained by means of the
local Lorentz invariance.

we choose vierbein $e_{\mu }^{a}$, spin connection $\omega _{abc}$ and the
matter field $\phi ^{A}$ as independent variables. Under the local Lorentz
transformation
\begin{equation}
e_{\mu }^{a}(x)\rightarrow e_{\mu }^{^{\prime }a}(x)=\Lambda
_{\;b}^{a}(x)e_{\mu }^{b}(x),\;\;\;\Lambda _{\;c}^{a}(x)\Lambda
_{\;b}^{c}(x)=\delta _{b}^{a},  \label{trans}
\end{equation}
$\omega _{abc}$ and $\phi ^{A}$ tranform as
\begin{equation}
\omega _{abc}(x)\rightarrow \omega _{abc}^{^{\prime }}(x)=\omega
_{lmn}(x)\Lambda _{\;a}^{l}\Lambda _{\;b}^{m}\Lambda _{\;c}^{n}+\Lambda
_{\;a}^{d}e_{d}^{\mu }\Lambda _{\;b}^{l}\partial _{\mu }\Lambda _{\;l}^{c}
\end{equation}
\begin{equation}
\phi ^{A}\rightarrow \phi ^{^{\prime }A}=D(\Lambda (x))_{\;B}^{A}\phi
^{B}(x).
\end{equation}
Since the coordinates $x^{\mu }$ \ do not transform under the local Lorentz
transformation, $\delta x^{\mu }=0$, from (\ref{6}), it can be proved that
in this case, $\delta _{0}\rightarrow \delta $. It is required that ${\cal L}%
_{m}$ is invariant under (\ref{trans}) and ${\cal L}_{g}$ is invariant
obviously. So under the local Lorentz transformation (\ref{trans}) ${\cal L}$
is invariant. In the light of the discussion in section 2, we would like to
have the relation
\begin{eqnarray}
& & \frac{\partial }{\partial x^{\mu }}(\frac{\partial {\cal
L}}{\partial
\partial _{\mu }e_{a}^{\nu }}\delta e_{a}^{\nu }+\frac{\partial {\cal L}}{%
\partial \partial _{\mu }\phi ^{A}}\delta \phi ^{A}+\frac{\partial {\cal L}}{%
\partial \partial _{\mu }\omega _{abc}}\delta \omega _{abc})
\nonumber \\
&+& [{\cal L]}_{e_{a}^{\nu }}\delta e_{a}^{\nu
}+[{\cal L]}_{\omega _{abc}}\delta \omega _{abc}+[{\cal L]}_{\phi
^{A}}\delta \phi ^{A}=0
\end{eqnarray}
where $[{\cal L]}_{e_{a}^{\nu }}$, $[{\cal L]}_{\omega _{abc}}$ and $[{\cal %
L]}_{\phi ^{A}}$ are the Euler expressions defined as
\[
\lbrack {\cal L}]_{e_{a}^{\nu }}=\frac{\partial {\cal L}}{\partial
e_{a}^{\nu }}-\partial _{\mu }\frac{\partial {\cal L}}{\partial \partial
_{\mu }e_{a}^{\nu }},
\]
\[
\lbrack {\cal L}]_{\omega _{abc}}=\frac{\partial {\cal L}}{\partial \omega
_{abc}}-\partial _{\mu }\frac{\partial {\cal L}}{\partial \partial _{\mu
}\omega _{abc}},
\]
\[
\lbrack {\cal L}]_{\phi ^{A}}=\frac{\partial {\cal L}}{\partial \phi ^{A}}%
-\partial _{\mu }\frac{\partial {\cal L}}{\partial \partial _{\mu }\phi ^{A}}%
.
\]
Using the equations $[{\cal L}]_{e_{a}^{\nu }}=0,$ $[{\cal L}]_{\omega
_{abc}}=0$ and $[{\cal L}]_{\phi ^{A}}=0$, we get the following expression
\begin{equation}
\frac{\partial }{\partial x^{\mu }}(\frac{\partial {\cal L}_{g}}{\partial
\partial _{\mu }e_{a}^{\nu }}\delta e_{a}^{\nu }+\frac{\partial {\cal L}_{g}%
}{\partial \partial _{\mu }\omega _{abc}}\delta \omega _{abc})+\frac{%
\partial }{\partial x^{\mu }}(\frac{\partial {\cal L}_{m}}{\partial \partial
_{\mu }e_{a}^{\nu }}\delta e_{a}^{\nu }+\frac{\partial {\cal L}_{m}}{%
\partial \partial _{\mu }\omega _{abc}}\delta \omega _{abc}+\frac{\partial
{\cal L}_{m}}{\partial \partial _{\mu }\phi ^{A}}\delta \phi ^{A})=0
\label{noether}
\end{equation}
where we have used the fact that only ${\cal L}_{m}$ contain the matter
field $\phi ^{A}$. Consider the infinitesimal local transformation $\Lambda
_{\;b}^{a}(x)=\delta _{b}^{a}+\alpha _{b}^{a}(x),$ $\alpha _{ab}=-\alpha
_{ba}$, $D(\Lambda )$ can be linearized as $[D(\Lambda )]_{\;B}^{A}=\delta
_{B}^{A}+\frac{1}{2}(I_{ab})_{\;B}^{A}\alpha _{ab}$, we have
\begin{equation}
\delta e_{a}^{\nu }=\alpha _{ab}e_{b}^{\nu }(x),
\end{equation}
\begin{equation}
\delta \omega _{abc}(x)=\alpha _{ad}\omega _{dbc}+\alpha _{bd}\omega
_{adc}+\alpha _{cd}\omega _{abd}+e_{a}^{\mu }\partial _{\mu }(\alpha _{bc})
\end{equation}
\begin{equation}
\delta \phi ^{A}=\frac{1}{2}(I_{ab})_{\;B}^{A}\phi ^{B}(x)\alpha _{ab}(x)
\end{equation}

We introduce $j_{ab}^{\mu }$%
\begin{eqnarray}
\sqrt{-g}j_{ab}^{\mu }\alpha _{ab} &=& \frac{3}{c}[\frac{\partial {\cal L}%
_{\omega }}{\partial \partial _{\mu }e_{a}^{\nu }}e_{b}^{\nu }\alpha _{ab}-%
\frac{\partial {\cal L}_{m}}{\partial \partial _{\mu }e_{a}^{\nu }}%
e_{b}^{\nu }\alpha _{ab} \nonumber \\
&-& \frac{\partial {\cal
L}_{m}}{\partial \partial _{\mu }\omega _{abc}}(\alpha _{ad}\omega
_{dbc}+\alpha _{bd}\omega _{adc}+\alpha _{cd}\omega
_{abd}+e_{a}^{\mu }\partial _{\mu }(\alpha
_{bc}))-\frac{1}{2}\frac{\partial {\cal L}_{m}}{\partial \partial
_{\mu }\phi ^{A}}(I_{ab})_{\;B}^{A}\phi ^{B}\alpha _{ab}],
\end{eqnarray}
then (\ref{noether}) can be rewritten as
\begin{equation}
\partial _{\mu }(\sqrt{-g}j_{ab}^{\mu }\alpha _{ab})+\frac{3c^{3}}{8\pi G}%
\partial _{\mu }(\frac{\partial {\cal L}_{\partial }}{\partial \partial
_{\mu }e_{a}^{\nu }}e_{b}^{\nu }\alpha _{ab}+\frac{\partial {\cal L}%
_{\partial }}{\partial \partial _{\mu }\omega _{abc}}\delta \omega _{abc})=0.
\label{222}
\end{equation}
From (\ref{total}) one can easily get that
\begin{equation}
\frac{\partial {\cal L}_{\partial }}{\partial \partial _{\mu }e_{a}^{\nu }}%
e_{b}^{\nu }\alpha _{ab}=\sqrt{-g}\omega _{a}e_{b}^{\mu }\alpha _{ab},
\end{equation}
\begin{equation}
\frac{\partial {\cal L}_{\partial }}{\partial \partial _{\mu }\omega _{abc}}%
\delta \omega _{abc}=\sqrt{-g}e_{b}^{\mu }(\omega _{a}\alpha
_{ba}+e_{c}^{\nu }\partial _{\nu }\alpha _{bc}),
\end{equation}
considering that $\alpha _{ab}$ is antisymmetry, i.e. $\alpha _{ab}=-\alpha
_{ba}$, we then get the result that
\begin{equation}
\frac{\partial {\cal L}_{\partial }}{\partial \partial _{\mu }e_{a}^{\nu }}%
e_{b}^{\nu }\alpha _{ab}+\frac{\partial {\cal L}_{\partial }}{\partial
\partial _{\mu }\omega _{abc}}\delta \omega _{abc}=\sqrt{-g}e_{a}^{\mu
}e_{b}^{\nu }\partial _{\nu }\alpha _{ab}.  \label{111}
\end{equation}
Defining a superpotential $V_{ab}^{\mu \nu }=e_{a}^{\mu }e_{b}^{\nu
}-e_{b}^{\mu }e_{a}^{\nu }$, and substituting (\ref{111}) into (\ref{222}),
we obtain
\begin{equation}
\partial _{\mu }(\sqrt{-g}j_{ab}^{\mu })\alpha _{ab}+[\sqrt{-g}j_{ab}^{\mu
}-(\frac{3c^{3}}{16\pi G})\partial _{\nu }(\sqrt{-g}V_{ab}^{\nu \mu
})]\partial _{\mu }\alpha _{ab}=0.  \label{ccc}
\end{equation}
Since $\alpha _{ab}$ and $\partial _{\mu }\alpha _{ab}$ are independent of
each other, we must have
\begin{equation}
\partial _{\mu }(\sqrt{-g}j_{ab}^{\mu })=0,  \label{j}
\end{equation}
\begin{equation}
j_{ab}^{\mu }=\frac{3c^{3}}{16\pi G}\frac{1}{\sqrt{-g}}\partial _{\nu }(%
\sqrt{-g}V_{ab}^{\nu \mu }).  \label{jj}
\end{equation}
From (\ref{j}) and (\ref{jj}), it can be concluded that $j_{ab}^{\mu }$ is
conserved identically. Sincer the current $j_{ab}^{\mu }$ is derived from
the local Lorentz invariance of the total Lagrangian, it can be interpreted
as the total angular momentum tensor density of the gravity-matter system.

For a globally hyperbolic Riemann-Cartan manifold, there exist Cauchy
surfaces $\Sigma _{t}$ foliating $M$. We choose a submanifold $D$ of $M$
joining any two Cauchy surfaces $\Sigma _{t_{1}}$ and $\Sigma _{t_{2}}$, so
the boundary $\partial D$ of $D$ consists of three parts: $\Sigma _{t_{1}}$,
$\Sigma _{t_{2}}$ and $A$ which is at spatial infinity. For an isolated
system, the space-time should be asymptotically flat at spatial infinity, so
the vierbein have the following asymptotical behaviour\cite{43,27,52}
\begin{equation}
\lim_{r\rightarrow \infty }(\partial _{\mu }e_{\nu a}-\partial _{\nu }e_{\mu
a})=0.  \label{631}
\end{equation}
Since
\[
\sqrt{-g}V_{ab}^{\mu \nu }=\frac{1}{2}\epsilon ^{\mu \nu \lambda \rho
}\epsilon _{abcd}e_{\lambda c}e_{\rho d},
\]
we have $\lim_{r\rightarrow \infty }\partial _{\lambda }(\sqrt{-g}%
V_{ab}^{\lambda \mu })=0$. Thus, we can get the total conservative angular
momentum from (\ref{j}) and (\ref{jj})
\[
J_{ab}=\int_{\Sigma _{t}}j_{ab}^{\mu }\sqrt{-g}d\Sigma _{\mu }=\frac{3c^{3}}{%
16\pi G}\int_{\partial \Sigma _{t}}\sqrt{-g}V_{ab}^{\mu \nu }d\sigma _{\mu
\nu },
\]
where $\sqrt{-g}d\Sigma _{\mu }$ is the covariant surface element of $\Sigma
_{t}$, $d\Sigma _{\mu }=\frac{1}{3!}\epsilon _{\mu \nu \lambda \rho }dx^{\nu
}\wedge dx^{\lambda }\wedge dx^{\rho }$, $d\sigma _{\mu \nu }=\frac{1}{2}%
\epsilon _{\mu \nu \lambda \rho }dx^{\lambda }\wedge dx^{\rho }$.

In summary, we have succeeded in obtaining an expression of an angular
momentum conservation law in Riemann-Cartan space-time. This conservation
las has the following main properties:

1. It is a covariant theory with respect to the generalized coordinate
transformations, but the angular momentum tensor is not covariant under the
local Lorentz transformation which, due to the equivalent principle, is
reasonable to require.

2. For a closed system, the total angular momentum does not depend on the
choice of the Riemannian coordinates and, according (\ref{631}), the
space-time at spatial infinity is flat, thus the conservative angular
momentum $J_{ab}$ should be a covariant object when we make a Lorentz
transformation $\Lambda _{ab}=A_{ab}=const.$ at spatial infinity, as in
special relativity
\[
J_{ab}^{^{\prime }}=A_{\;a}^{c}A_{\;b}^{d}J_{cd}.
\]
To understand this, the key point is that to obtain $J_{ab}$, one has to
enclose everything of the closed system, and every point of space-time at
spatial infinity belongs to the same Minkowski space-time in that region.
This means that in general relativity for a closed system, the total angular
momentum $J_{ab}$ must be looked upon as a Lorentz tensor like that in
special relativity.

\section*{Acknowledgment}

The author gratefully acknowledges the support of K. C. Wong Education
Foundation, Hong Kong.

\end{document}